\nonstopmode

\documentclass[aps,prl,twocolumn,showpacs]{revtex4}
\usepackage{graphicx}
\usepackage{float}

\begin{document}
\title{Straining graphene by chemical vapor deposition growth on copper}
\author{V. Yu}
\author{E. Whiteway}
\author{J. Maassen}
\author{M. Hilke}
\affiliation{Department of Physics, McGill University, Montr\'eal, Canada, H3A 2T8}
\date{\today}

\begin{abstract}
Strain can be used as an alternate way to tune the electronic properties of graphene. Here we demonstrate that it is possible to tune the uniform strain of graphene simply by changing the chemical vapor deposition growth temperature of graphene on copper. Due to the cooling of the graphene on copper system, we can induce a uniform compressive strain on graphene. The strain is analyzed by Raman spectroscopy, where a shift in the 2D peak is observed and compared to our {\it ab initio} calculations of the graphene on copper system as a function of strain.
\end{abstract}
\maketitle

The discovery of the electric field effect in a few layers of graphene \cite{Novo04,deHeer04} led to a genuine explosion of work on graphene with the subsequent observation of the anomalous quantum Hall effect in graphene monolayers \cite{Novo05,Kim05} following its theoretical prediction \cite{Ando02}. The most important property for electronic device applications is the possibility to tune the resistance and electronic density of the graphene sheet. This is achieved by using the substrate directly as a backgate \cite{Novo04,Novo05,Kim05} or by processing a top gate \cite{deHeer04}.

More recently, it was suggested theoretically that it is possible to tune the electronic density of graphene and create confinement \cite{Pereira09}, as well as to observe a zero-field quantum Hall effect by using strain engineering \cite{Guinea09}. In general, strain as a way to enhance electronic and optical properties in semiconductors has been very successful \cite{strain}. Mechanically, graphene turns out to be of very high strength \cite{strength} and is stretchable up to 30\% in some cases \cite{kim09}.

While exfoliated graphene bears very high mobilities when suspended \cite{highmu}, the size is limited to a few microns. Large scale epitaxial graphene on SiC is an interesting avenue but suffers from high substrate costs and difficulty to transfer \cite{SiC}. Alternatively, large scale graphene can be synthesized by chemical vapor deposition (CVD) using a polycrystalline metal foil as a catalyst \cite{kim09,reina08}.

\begin{figure}[!h]
\centering
\vspace{0cm}
\includegraphics[width=7cm]{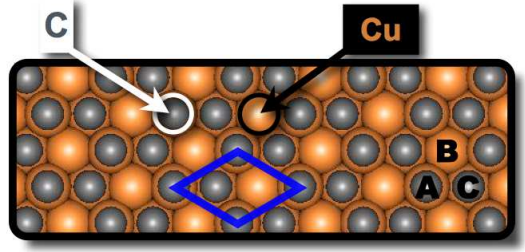}
\vspace{-0.3cm}
\caption{View of graphene on a Cu(111) surface. The grey and dark orange spheres represent the C and Cu atoms, respectively. The blue parallelogram delimits the area of the supercell used for the first principles calculations. The letters A, B and C depict the different Cu(111) sites. The minimal energy configuration of graphene on copper corresponds to one sub-lattice carbon sitting at the top-site (A-site) and the other sub-lattice carbon residing at the hollow-site (C-site).}
\label{picture}
\end{figure}

In this letter, we demonstrate that it is possible to strain graphene intrinsically and uniformly during the graphene growth process. This leads to interesting new properties, such as a change in Fermi energy and Fermi velocity in graphene. This opens new doors to the possibility of engineering devices, where the electronic properties are tuned by strain. We consider the situation depicted in figure \ref{picture}, where a graphene monolayer is sitting on a Cu surface. For our {\it ab initio} calculations we consider a Cu(111) surface, where the graphene and the Cu share the same in-plane lattice constant \cite{dft}. We assume that the system has no structural defects, meaning that the graphene and the Cu match perfectly at the interface. In this case, the most stable configuration of graphene on Cu(111) corresponds to one carbon sitting at the top-site (A-site) and the other carbon located at the hollow-site (C-site) \cite{gio,maassen1}. The same configuration also applies to graphene on other metallic surfaces \cite{maassen2}. To induce strain, we uniformly expand or compress the in-plane lattice constant of the whole graphene/Cu system.

Zero strain corresponds to a C-C bond length of $a_{eq}$=1.415\,\AA, which was found to minimize the total energy of an isolated graphene sheet using density functional theory. For every $a$, we (i) perform a structural relaxation of the system to ensure that the forces acting on each atom is less than 0.01\,eV/\AA, (ii) obtain the self-consistent electronic density and Hamiltonian and (iii) calculate the band structure of the hybrid graphene/Cu system along the high-symmetry points $\rm \Gamma$, M and K.

\begin{figure}[!h]
\centering
\vspace{2cm}
\includegraphics[width=8cm]{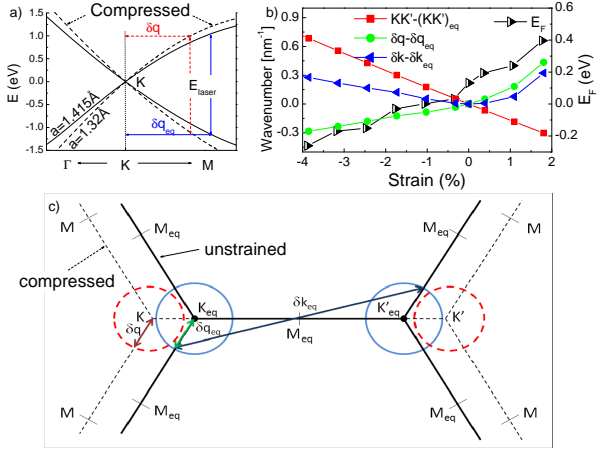}
\vspace{-2cm}
\caption{a) Band structure of graphene on Cu close to the K point for two different values of strain, i.e., 0\% (solid line) and -6.7\% (dashed line). b) depicts the change of $E_F$ (relative to the Dirac point), $\delta k$, $\delta q$ and ${\rm KK}'$ as a function of strain. c) Partial view of the Brillouin zone. The arrow labeled $\delta k_{eq}$ indicates the phonon process at the origin of the unstrained 2D Raman peak. The arrow labeled $\delta q_{eq}$ corresponds to the wavenumber of the Raman excited electron of the unstrained lattice and $\delta q$ for the strained (compressed) lattice. The circles show the constant energy surface of the Raman excited electron in the unstrained (solid line) and strained (dashed lined) lattice.}
\label{modelling}
\end{figure}

The most striking features induced by the strain are the change in Fermi velocity close to the Dirac point and more notably a shift in the Fermi energy with respect to the Dirac point as a function of strain as shown in figure \ref{modelling} b). This is very interesting, since this allows for the tuning of the electronic density in graphene by strain.

Experimentally, we expect to be able to strain graphene on Cu simply by heating the combined system, since the thermal expansion of Cu is different from that of graphene. While graphene has a small negative thermal expansion coefficient $\alpha$ \cite{Bao09}, for Cu we have $\alpha\simeq 25.8\times 10^{-6}$ $/^{\circ}$C \cite{toulouk70} at 1000$^\circ$C. This would lead to a relative graphene compression of 0.5\% from 1100 to 900$^{\circ}$C.

We synthesized graphene monolayers by CVD of hydrocarbons on 25 $\mu$m-thick commercial polycrystalline copper foils. The Cu foil is first acid-treated for 10 mins using acetic acid and then washed thoroughly with de-ionized water. Graphene growth is realized in similar conditions to the recently reported ones \cite{li09,bae10}, but using a vertical quartz tube. Graphene is grown with temperatures ranging from 900$^\circ$C to 1100$^\circ$C in 0.5 Torr, with a 4 sccm H$_2$ flow and a 40 sccm CH$_4$ flow for 30 minutes.  During the cooling process the methane flow is stopped while the hydrogen flow is kept on.

\begin{figure}[!h]
\centering
\includegraphics[width=8cm]{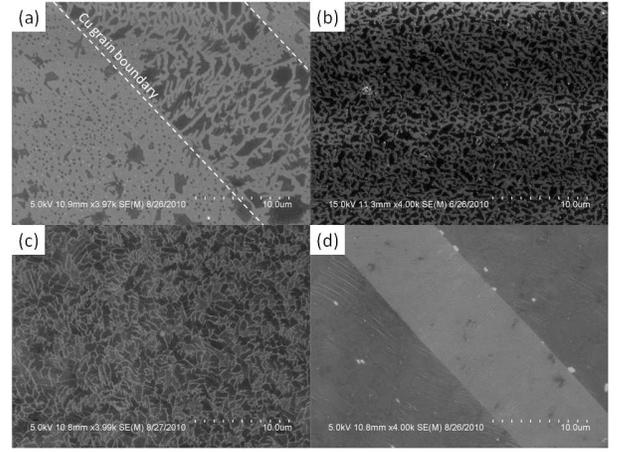}
\caption{SEM images of graphene on Cu grown for 30 minutes with different growth temperatures: (a) 900$^\circ$C, (b) 950$^\circ$C (c) 975$^\circ$C, and (d) 1050$^\circ$C, respectively. Each image is approximately 30$\mu$m$\times 30 \mu$m.}
\label{sem_image}
\end{figure}

Figure \ref{sem_image} illustrates the temperature effect on the synthesis of graphene. These scanning electron microscope (SEM) images show that the density and size of growth domains increase with temperature. At the lowest growth temperatures we did not observe full coverage, even for longer growth times. 

\begin{figure}[!h]
\centering
\includegraphics[width=8cm]{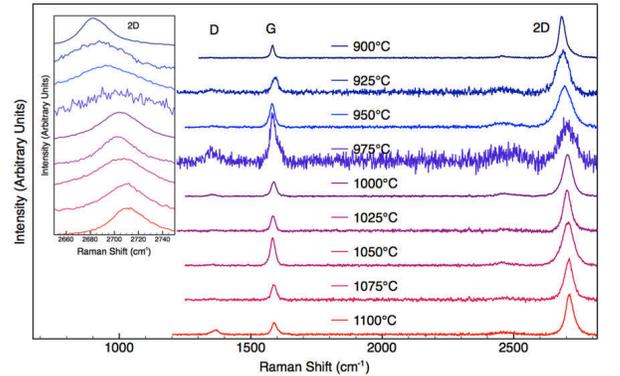}
\caption{Raman spectra for each growth temperature normalized to the 2D peak. Inset: Close-up of the 2D peaks showing the shifts of the peaks. The 975$^\circ$C spectrum shows a region of possible bilayer.}
\label{spectral}
\end{figure}

In order to characterize the graphene layer further, we used Raman spectroscopy. Raman spectra are measured with a 100x objective at 514nm, having a 1800 grooves/mm grating and spectral resolution of about 2 cm$^{-1}$. The results are shown in figure \ref{spectral}. There are two dominant peaks, labeled G and 2D. The G peak around 1576 cm$^{-1}$ arises from a first order G band phonon process. The energy corresponds to the phonon energy at the $\Gamma$ point, which is degenerate. Several authors have shown that the degeneracy is lifted by the application of uniaxial strain \cite{Huang09}, which leads to a splitting of the G peak. For uniform biaxial strain, there are no experiments on compressive strain in graphene, but we expect the strain to simply renormalize the phonon energy ($\omega^G$) following $\omega^G=\omega^G_{eq}(1-2\epsilon\gamma)$ \cite{malard09,Thomson02}. $\gamma\simeq 1.24$ is the Gr\"uneisen parameter in carbon nanotubes \cite{Reich00}, $\epsilon=(a-a_{eq})/a_{eq}$ is the strain and the factor 2 stems from the biaxial strain in both directions.

The other dominant peak is labeled 2D (also called G$'$) and stems from a two phonon process. The two phonons of wavenumber $\delta k$ and -$\delta k$ scatter the electron with wavenumber $\delta q$ into the other valley (K to K$'$) and back as illustrated in figure \ref{modelling} c). The phonon energy depends on the dispersion close to the Dirac points (K or K$'$) along the M direction. Hence $\delta k$ will depend on the electronic band structure, which is shown in figure \ref{modelling} a) for two different values of strain. Using the calculated electronic band structure we can extract $\delta q$ in the M direction for our Raman laser energy $E_{laser}=2.412$\,eV (514\,nm). From a geometrical consideration, we have
$\delta k  =  \sqrt{({\rm KK}')^2 + 2\cdot {\rm KK}' \cdot \delta q + 4\cdot \delta q^2}$ (see figure \ref{modelling} c)), where KK$'$=K$-$K$'$ is the distance in reciprocal space between the two valleys K and K$'$. KK$'$ will simply be renormalized by a uniform strain leading to ${\rm KK}'=({\rm KK}')_{eq} / (1+\epsilon)$, which is shown in figure \ref{modelling} b). For small values of $\epsilon$, the renormalization constant is approximately linear in strain. $\delta k$, on the other hand, is non-linear as it depends on $\delta q$.

\begin{figure}[!h]
\centering
\includegraphics[width=8cm]{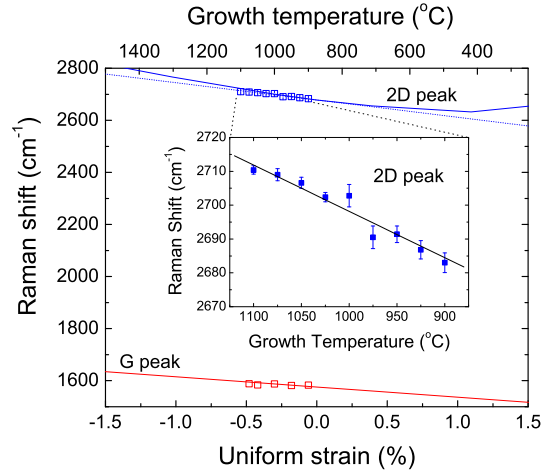}
\caption{The Raman shift of the 2D and G peaks as a function of strain obtained from the {\it ab initio} results. The solid line and the dotted line for the 2D peak are obtained with and without considering the effect of the electronic band structure. The square symbols are the experimental data points obtained for different growth temperatures. The inset shows a zoom-in of the 2D peak data points and the line represents the best linear fit.}
\label{strain}
\end{figure}

In this region, the phonon band velocity can be parameterized by $v_{iTO}=5.47\times10^{-3}v_F$ \cite{malard09}, where $v_F\simeq 10^{6}$m/s is the Fermi velocity. In order to evaluate the change in phonon energy due to uniform strain for the 2D peak we simply combine the change in wavenumber $\delta$k from the linear dispersion with the renormalized total phonon energy due to the Gr\"uneisen parameter as $\omega^{2D}\simeq[\omega^{2D}_{eq}+2v_{iTO}(\delta k-\delta k_{eq})](1-2\epsilon\gamma)$. The factor 2 in front of the band velocity term stems from the phonon pair involved in the process. For the equilibrium (unstrained) values of $\omega^{(2D,G)}_{eq}$, we use those obtained for exfoliated graphene at the same wavelength (514nm) \cite{yu09}, i.e., $\omega^{2D}_{eq}=2678$cm$^{-1}$ and $\omega^{G}_{eq}=1576$cm$^{-1}$. We can now evaluate the expected frequencies for the 2D and G peaks as a function of a uniform strain, which is shown in figure \ref{strain}. The corresponding growth temperature dependence of the 2D peak data is then fitted to the 2D peak position obtained from the strain calculation, assuming a linear dependence. We also show the result (in dotted lines) for the 2D peak computed without the band structure effect, which is then simply a linear function of strain. By including the band structure effect, the 2D peak dependence becomes non-linear with strain. This is in contrast to the G peak, which does not depend on the electronic band structure.

The data fits nicely with the expected Raman shift when the difference in strain between growth temperatures of 900 and 1100$^\circ$C is -0.5\%, which is the expected value from the thermal expansion of copper for this temperature range. This clearly demonstrates that it is possible to strain graphene uniformly, simply by changing the CVD growth temperature, which is one of the important results of this letter. This also shows that when graphene is grown at high temperatures it is under small compressive strain at room temperature (the Raman spectra are taken at room temperature).

The peak labeled D in figure \ref{spectral} is located at about half the frequency of the 2D peak, since it involves only a one phonon process. This process, however, is only activated in the presence of disorder \cite{casiraghi09,malard09}. Raman spectroscopy can be employed to determine the number of graphene layers \cite{ferrari06}, which confirms that we have mainly single layered graphene as expected in CVD on Cu \cite{li09}.

It is instructive to compare our results to those obtained for exfoliated graphene, where several groups have been able to detect uniaxial stain in exfoliated graphene flakes using Raman spectroscopy \cite{ni08}. The reported shifts in the 2D peak vary from  -7.8 cm$^{-1}$/1\% to -66 cm$^{-1}$/1\%. It was recently suggested that the large range of values could be due to the dependence on the polarization of the laser beam and orientation of the graphene lattice \cite{huang10}. Our results, between 0 and -1\% strain, give values of -66 cm$^{-1}$/1\% (without band structure effect) and -79 cm$^{-1}$/1\% (with band structure effect), which is the higher bound of the reported values for uniaxial strain. This is consistent with expectations, since for biaxial strain we expect approximately a doubling of the strain induced Raman shift. In a very different geometry, where stretched biaxial strain has been measured by depositing a graphene flake on a shallow depression, a very high 2D peak shift of about 200 cm$^{-1}$/1\% was reported \cite{metzger10}. In epitaxial graphene, shifts of the principal Raman peaks were observed to be dependent on the growth time for a fixed annealing temperature \cite{ferralis08}. For CVD on nickel, no systematic shift with growth temperature was observed \cite{chae09}.

The formed graphene/Cu heterostructure constitutes a very interesting system by itself. Not only can the amount of relative strain be tuned as shown above, leading to changes in the Fermi energy and Fermi velocity, but this system also opens the door to numerous applications. For instance, the added graphene layer can enhance electronic properties in Cu interconnects, for example, or greatly enhance thermal conductivities of thin Cu films. The tunability of the strain of the graphene layer, greatly increases this potential. The graphene/Cu heterostructure can also be selectively etched for partial or full transfer onto other substrates, which can lead to interesting transport properties such as a very sharp weak localization peak \cite{Whiteway10}.

We would like to acknowledge help from S. Elouatik for the Raman characterization and the use of MIAM microfab facilities and GCM at Polytechnique for processing and characterization in addition to financial support from RQMP, NSERC and FQRNT.

\end{document}